\documentstyle[preprint,aps]{revtex}
\begin{document}
\draft
\preprint{\vbox{\hbox{CLNS 96/1408}\hbox{LBL-38656}\hbox{UCB-96/17}}}
\title{Perturbative Scattering Phase Shifts in One-Dimension: Closed-form
results}
\author{C.K. Au}
\address{Department of Physics and Astronomy, University of South Carolina,
Columbia, SC 29208.}
\author{Chi--Keung Chow}
\address{Newman Laboratory of Nuclear Studies, Cornell University, Ithaca,
NY 14853.}
\author{Chong--Sun Chu}
\address{Department of Physics and Theoretical Physics Group, \\
Lawrence Berkeley Laboratory, University of California, Berkeley, CA 94720.}
\date{\today}
\maketitle
\begin{abstract}
A simple closed form expression is obtained for the scattering phase shift
perturbatively to any given order in effective one-dimensional problems.
The result is a hierarchical scheme, expressible in quadratures, requiring
only knowledge of the zeroth order solution and the perturbation potential.
\end{abstract}
\narrowtext
Since most quantum mechanical problems cannot be solved exactly, perturbation
theory is a useful tool.
Standard perturbation theory relies on the use of the Green function or the
spectral summation over the intermediate states.
Variants of the standard perturbation theory were introduced by Schrodinger
\cite{1}, Podolsky \cite{2}, Sternheimer \cite{3} and Dalgarno and Lewis
\cite{4}.
Originally, the Dalgarno--Lewis method was introduced as a means of
calculating sum rules.
Another useful variation of perturbation theory was introduced by Wentzel
\cite{5} and later by Price \cite{6} and by Polikanov \cite{7}.
However, this method was overlooked until its rediscovery by Aharonov and
Au \cite{8} and more or less about the same time by Turbiner \cite{9}.
This method has now come to be widely known as ``Logarithmic Perturbation
Theory'' or LPT for short.

LPT has many advantages.
In one-dimensional problems, it becomes possible to obtain perturbative
solutions to all order in a hierarchical scheme, for the energies and wave
functions for the bound states \cite{7,8,9} and for the phase shifts in
scattering states \cite{10}.
A few modifications of LPT to handle the presence of zeros in the wave
functions in excited bound states were given by Au {\it et. al.} \cite{11}.

In three-dimensions, the zeros of excited bound states appear as nodal
surfaces.
A variation of LPT to address this difficulty was given by Au \cite{12}.
In this method, the bound state wave function is written as $F\exp(-G)$,
and a perturbation expansion is carried out only on $F$.
In the absence of any nodes, the zeroth order function $F_0$ can be set equal
to unity as the information on the unperturbed wave function can be totally
absorbed into $G$.
This is then equivalent to writing the perturbed wave function as a scalar
function times the unperturbed wave function.
When confined only to the first order correction, this method yields the
equivalent of the Dalgarno--Lewis method.
Henceforth, we refer this method of writing the perturbed wave function as a
product of the unperturbed wave function and a scalar function as the
Dalgarno--Lewis Perturbation Theory (DLPT for short).
The difficulties associated with the presence of zeros in the wave function
when using DLPT are also discussed in reference \cite{12}.

In Eq. (1.25) of Ref. \cite{12}, the hierarchical structure of DLPT
is displayed explicitly for a multidimensional system.
For one-dimensional system, this hierarchy can be trivially integrated twice
to give the perturbative solutions in quadrature to reproduce the
results of Kim and Sukhatme \cite{13}.
Recently, Nandi {\it et. al.} \cite{14} applied DLPT to the scattering phase
shifts in one-dimension and rederived the results of Au {\it et. al.}
\cite{10} obtained via LPT.

The equivalence of DLPT and LPT was first discussed by Au and Aharonov
\cite{15}, and subsequently by many other authors \cite{16}.
In the absence of nodal difficulties, this equivalence can be expected simply
from the well defined mapping between the wave function and its logarithm.
>From the Wronskian conditions, one can easily see that the scattering wave
function in one-dimension with the expected asymptotic behavior is free from
zeros and hence its logarithm is regular.
It follows from the above that LPT and DLPT should produce the same
perturbative scattering phase shifts.

In this letter, we apply DLPT to the scattering problem in one-dimension and
derive a closed form expression for the perturbative correction to the phase
shift to any order.
This is an improvement over earlier works where the calculational procedures
were laid out, but no such closed form results were available.

The unperturbed wave function $\psi_0(x)$ is given by the Schroedinger
equation
\begin{equation}
- \textstyle{1\over2} \psi_0''(x) + V(x) \psi_0(x) = \textstyle{1\over2} k^2
\psi_0(x),
\end{equation}
for a particle with energy ${1\over2}k^2$.
In our notation $\hbar=m=1$.
Similarly, $\psi(x)$, the perturbed wave function, is given by
\begin{equation}
- \textstyle{1\over2} \psi''(x) + V(x) \psi(x) + \lambda U(x) \psi(x)
= \textstyle{1\over2} k^2 \psi(x),
\end{equation}
with $\lambda U(x)$ the perturbing potential.
Both Schrodinger equations are subjected to the boundary condition at
infinity.
\begin{equation}
\psi_0(x) = \psi(x) = \exp(-ikx),\quad\hbox{ as}\quad x\to\infty.
\end{equation}
Then the phase shifts are defined as their respective phases at $x=0$.
\begin{equation}
\exp(-2i\delta_0) = \psi_0^*(0)/\psi_0(0), \quad
\exp(-2i\delta) = \psi^*(0)/\psi(0).
\label{ps}
\end{equation}
The Wronskian condition,
\begin{equation}
\psi_0(x){\psi_0^*}'(x)-\psi^*_0(x)\psi'_0(x) = 2ik,
\label{wro}
\end{equation}
prevents      $\psi_0(x)$ from having any node, and
ensures that the ratio
\begin{equation}
f(x) = \psi(x)/\psi_0(x)
\end{equation}
is well-defined.
The perturbative correction, in terms of $f(x)$, is given by
\begin{equation}
\psi_0(x)f''(x) + \lambda \psi'_0(x)f'(x) - 2 \lambda U(x) \psi_0(x)f(x) = 0.
\label{sch}
\end{equation}
where one can perform a perturbative expansion on $f(x)$,
\begin{equation}
f(x)=f_0(x) + \lambda f_1(x) + \lambda^2 f_2(x) + \dots,
\end{equation}
with $f_0(x)\equiv 1$ identically.
\begin{mathletters}
Performing this expansion on Eq. (\ref{sch}), one gets to zeroth order,
\begin{eqnarray}
\lambda^0: &\quad \psi_0(x)f''_0(x) + 2 \psi'_0(x)f'_0(x) &=0 \\
\noalign{{\rm \noindent and to the nth order,}}
\lambda^n: &\quad \psi_0(x)f''_n(x) + 2 \psi'_0(x)f'_n(x) &= 2U(x)\psi_0(x)
f_{n-1}(x),\quad n\geq1.
\end{eqnarray}
\end{mathletters}
The first one vanishes identically as $f_0(x)\equiv 1$, while the second one
can be recasted in the form
\begin{equation}
(\psi_0^2(x) f'_n(x))' = 2 U(x) \psi_0^2(x) f_{n-1}(x),
\end{equation}
which can be integrated from infinity to give in a hierarchical scheme:
\begin{equation}
f_n(x) = \int_x^\infty dy\, {1\over\psi_0^2(y)} \int_y^\infty dz\, 2 U(z) \,
\psi_0^2(z)\, f_{n-1}(z).
\label{2i}
\end{equation}
As a result, one can apply it recursively to obtain $f_n(x)$ of arbitrarily
high order.

Relation (\ref{2i}) can be solved in terms of the zeroth order information.
The Wronskian condition (\ref{wro}) can be rewritten as
\begin{equation}
\psi_0^2(x) \left({\psi_0^*(x)\over\psi_0(x)}\right)' = \rho(x) q'(x) = 2ik,
\end{equation}
where the definitions of $\rho(x)$ and $q(x)$ are identical to that in Ref.
\cite{10,11}.
\begin{equation}
\rho(x) = \psi_0^2(x), \quad
q(x) = {\psi_0^*(x)\over\psi_0(x)} - {\psi_0^*(0)\over\psi_0(0)}.
\end{equation}
Then relation (\ref{2i}) becomes, upon using these two relations:
\begin{eqnarray}
f_n(x)&=& \int_x^\infty dz\, 2 U(z) \,\rho(z)\, f_{n-1}(z) \int_x^z dy\,
{1\over\rho(y)} \nonumber\\ &=& {1\over ik}
\int_x^\infty dz\, 2 U(z) \,\rho(z)\, f_{n-1}(z)\, (q(z) - q(x))
= J[f_{n-1}](x),
\label{1i}
\end{eqnarray}
with
\begin{equation}
J[g](x) = {1\over ik} \int_x^\infty dz\, 2 U(z) \,\rho(z)\,
(q(z) - q(x)) \, g(z).
\end{equation}
So one can apply Eq. (\ref{1i}) iteratively and gets
\begin{equation}
f_n(x) =\overbrace{J[J[\cdots J[}^{n\rm\;times}1]\cdots ]](x).
\label{f}
\end{equation}

After obtaining the expressions for $f_n(x)$, the only remaining task is to
express $\delta_n$ in terms of these $f_n$'s.
Recall that the phase shift corrections $\delta_n$ are naturally defined by
\begin{equation}
\delta = \delta_0 + \lambda \delta_1 + \lambda^2 \delta_2 + \cdots .
\end{equation}
>From the definitions of $f(x)$ and $\delta$, it is easy to see that
\begin{equation}
\delta - \delta_0 = {\rm Im} \log f(0).
\end{equation}
Performing an expansion in powers of $\lambda$, one can express $\delta_n$
as a function of the $n$-th derivative of $\log f(0)$,
\begin{equation}
\delta_n = {\rm Im} {1\over n!} \left( {d^n\over d\lambda^n} \log f(0) \right)
_{\lambda=0} .
\label{dn}
\end{equation}

Now we have obtained the correction of $f(x)$ to arbitrary order by Eq.
(\ref{f}), and expressed $\delta_n$ in terms of $f(0)$ by Eq. (\ref{dn}).
One only need to combine these two pieces of knowledge to get a closed form
expression of $\delta_n$.
This can be most conveniently achieved by the lemma below.

Lemma:  If $g=g(f(\lambda))$ and $f(\lambda)=1+\lambda f_1+\lambda^2 f_2
+ \dots$, then
\begin{equation}
{1\over n!} {d^ng\over d\lambda^n}\Bigg|_{\lambda=0} = \sum_{\{i_p\}_n}
{1\over i_1! i_2! \cdots i_n!} g^{(j)} f_1^{i_1} f_2^{i_2} \dots f_n^{i_n},
\end{equation}
where $\{i_p\}_n=\{(i_1,i_2,\dots,i_n)\}$ is the set of $n$ non-negative
integers satisfying 
\begin{equation}
\sum_{p=1}^n pi_p = n,
\end{equation}
$j$ is defined by 
$j=\sum_{p=1}^n i_p$,
and ${\displaystyle g^{(j)}={d^jg\over df^j}\Big|_{f=1}}$.

The proof is straightforward, just perform the double Taylor expansion and
collect terms of the same order in $\lambda$.
For our purpose, $g=\log f$ and hence $g^{(j)} = (-1)^{j-1} (j-1)!$.
This leads us to the central result of this paper,
\begin{equation}
\delta_n = {\rm Im} \sum_{\{i_p\}_n} (-1)^{j-1} {(j-1)!\over i_1! i_2! \dots
i_n!} f_1^{i_1} f_2^{i_2} \dots f_n^{i_n},
\label{cfe}
\end{equation}
where $f_n\equiv f_n(0)$ is given in Eq. (\ref{f}).

Physical results should be independent of which perturbative scheme one
chooses to work with, and the $\delta_n$ obtained above should agree with
the standard Rayleigh--Schrodinger results as well as those from LPT.
The agreement of LPT with the Rayleigh--Schrodinger theory on phase shifts was
demonstrated explicitly in Ref. \cite{10} up to order $\lambda^4$.
We shall demonstrate below that Eq. (\ref{cfe}) also reproduces exactly the 
same expressions for $\delta_n$ given by LPT.

For any given $n$, we are going to enumerate all the non-negative integer
$n$-plet $(i_1, i_2,\dots,i_n)$ which satisfy the condition
$\sum_{p=1}^n pi_p = n$.
Each of these $n$-plet is going to give a term in Eq. (\ref{cfe}).
The integral expressions for $f_n(0)$'s are substituted in, and the rest are
just simplifications.

For $n=1$, there is only one element in $\{i_p\}$, namely $(1)$.
Then Eq. (\ref{cfe}) gives
\begin{eqnarray}
\delta_1 = {\rm Im}\, f_1
&=& {\rm Im}\,{1\over ik} \int_0^\infty dx\, U(x)\, \rho(x)\, q(x) \nonumber\\
&=& {-1\over k} {\rm Re} \int_0^\infty dx\, U(x)\, \rho(x)\, q(x).
\end{eqnarray}
Note that $q(0) = 0$.
This agrees with the LPT result (Eq. (26) in Ref. \cite{10}).
In the notation defined in Ref. \cite{10}, with
\begin{equation}
I[F_1, F_2, \dots, F_n] = \int_0^\infty dx_1\, F_1(x_1)\,
\int_{x_1}^\infty dx_2\, F_2(x_2)\, \dots \int_{x_{n-1}}^\infty dx_n F_n(x_n),
\end{equation}
the result is
\begin{equation}
\delta_1 = {-1\over k} {\rm Re}\, I[U\rho q].
\end{equation}
This notation will facilitate the presentation of higher order results.

Less trivial is the case $n=2$, where $(i_1, i_2)$ can be either $(0,1)$ or
$(2,0)$.
Eq. (21) then gives
\begin{equation}
\delta_2 = {\rm Im} (f_2 - \textstyle{1\over2} f_1^2).
\end{equation}
In particular,
\begin{eqnarray}
f_2 &=& {-1\over k^2} \int_0^\infty dx\, U(x)\, \rho(x)\, q(x)\,
\int_x^\infty dy\, U(y)\, \rho(y)\, (q(y) - q(x)) \nonumber\\
&=& {1\over k^2} I[U\rho q^2, U\rho] + {-1\over k^2}  I[U\rho q, U\rho q]
\nonumber\\ &=& {1\over k^2} I[U\rho q^2, U\rho] + {1\over2}f_1^2.
\end{eqnarray}
As a result, we have reproduced Eq. (31) in Ref. \cite{10}:
\begin{equation}
\delta_2 = {1\over k^2} {\rm Im}\, I[U\rho q^2, U\rho].
\end{equation}

The $n=3$ is more cumbersome but still straightforward.
In this case $(i_1, i_2, i_3)$ can be $(0,0,1)$, $(1,1,0)$ and $(3,0,0)$.
Hence
\begin{equation}
\delta_3 = {\rm Im} (f_3 - f_1 f_2 + \textstyle{1\over3} f_1^3).
\end{equation}
After expanding out the $f_n$'s and some changes of variables, the result can
be casted into the form
\begin{equation}
\delta_3 = {2\over k^3} {\rm Re}\, I[U\rho q^2, U\rho q, U\rho],
\end{equation}
which has been reported in Eq. (33) of Ref. \cite{10}.

Similar exercises can be done with even higher $n$.
We do not think, however, that it is instructive to present those higher
order calculations in full details.
We just want to report that the $\delta_4$ expression also agrees with that
from LPT.
The possible $(i_1,i_2,i_3,i_4)$'s are $(0,0,0,1)$, $(1,0,1,0)$, $(0,2,0,0)$,
$(2,1,0,0)$ and $(4,0,0,0)$.
Note that the number of elements in $\{i_p\}_n$ is increasing faster than
$n$.

The central result of this article is Eq. (\ref{cfe}), which is a closed
form expression for $\delta_n$, obtained for the first time, and
expressible in terms of only zeroth order and on-shell informations.
We have also explicitly checked that the results for $n\leq 4$ agree with
those from LPT.
Admittedly, the results are still not as simple as one would like to see.
There are two sources of complexity of the results.
Firstly, the cardinality of $\{i_p\}_n$ is growing with $n$.
It is easy to see that combinatorically this is the same as the number of
inequivalent irreducible representations of the symmetric group $S_n$ 
\cite{tung}.
Graphically, this is equal to the number of ways of drawing different Young 
diagrams with $n$ boxes.
The other complication is that each $f_n$, expressed as iterated
$J[\>\cdot\>]$'s, breaks down to many terms of the form $I[\>\cdot\>]$'s.
As a result, we expect the higher order results to grow more and more messy.

Still, this drawback is compensated by another nice feature of the scheme,
namely the freedom to choose the unperturbed system.
Since only on-shell information is needed in our scheme, it is much easier
to find a ``unperturbed'' state which approximates the exact system than in
the case of the Rayleigh--Schrodinger scheme, in which off-shell information
is also needed and one needs to solve the ``unperturbed'' system completely.
As discussed in Ref. \cite{10}, one expects $\delta_n$ converges quickly as
long as ``Levinson criterion'' (the unperturbed and perturbed system should
have the same number of bound states) is satisfied.
Hence the first several terms in the perturbation series should suffice to
give a good approximation of the exact $\delta$, and the higher order terms
are seldom needed in practice.

In conclusion, we have obtained the perturbative correction to the scattering 
phase shift of any effectively one-dimensional Schrodinger problem.  
The result can be expressed in closed form and only on-shell information is 
needed.  
With a suitable choice of the ``unperturbed'' system, the perturbation series 
converges rapidly.  
We have also demonstrated explicitly that the our results agree with the 
conventional LPT results up to the fourth order in the expansion series.  

\acknowledgments
The work of C.S.C. was supported in part by the Director, Office of
Energy Research, Office of High Energy and Nuclear Physics, Division of
High Energy Physics of the U.S. Department of Energy under Contract
DE-AC03-76SF00098 and in part by the National Science Foundation under
grant PHY-9514797.  
The work of C.K.C. is also supported in part by the National Science 
Foundation.


\begin{references}
\bibitem{1} Schrodinger E 1926 Ann. Phys. (Paris) 81, 112; 1926 ibid 79, 745.
\bibitem{2} Podosky B 1928 Proc. Natl. Acad. Sci. U.S. A. 14 253.
\bibitem{3} Sternheimer R 1951 Phys. Rev. 84 244.
\bibitem{4} Dalgarno A and Lewis J T 1955 Proc. Roy. Soc. London 233 70.
\bibitem{5} Wentzel G 1926 Z. Phys. 38 518.
\bibitem{6} Price R J 1954 Proc. Phys. Soc. 67 383.
\bibitem{7} Polikanov V S 1967 Zh. Eksp. Teor. Fiz. 52 1326 (1967 Sov. Phys.-
    JETP 25, 882); 1975 Theor. Math. Phys. (USSR) 24 230.
\bibitem{8} Aharonov Y and Au C K 1979 Phys. Rev. Lett. 42 1582.
\bibitem{9} Turbiner A V Zh. 1980 Zh. Eksp. Teor. Fiz. 79 1719 (1980 Sov.
    Phys.-JETP 52, 868).
\bibitem{10} Au CK, Chow C K, Chu C S, Leung P T and Young K 1992 Phys. Lett.
     A164 23.
\bibitem{11} Au C K, Chan K L, Chow C K, Chu C S and Young K 1991 J Phys.
     A24, 3837.
\bibitem{12} Au C K 1984 Phys. Rev A29 1034.
\bibitem{13} Kim I-W and Sukhatme U P 1992 J Phys. A25, L647.
\bibitem{14} Nandi T K, Bera P K, Panja M M and Talukar B 1996 J Phys A
     29, 1101.
\bibitem{15} Au C K and Aharonov Y 1979 Phys. Rev. A20 2245.
\bibitem{16} See for example, Mavromatis H A 1991 Am J Phys. 59 738
\bibitem{tung} See for example, Tung W K, Group Theory in Physics,
     World Sci., 1985.
\end{references}
\end{document}